# 3D Nanostructures on Ge/Si(100) Wetting Layers: Hillocks and Pre-Quantum Dots

Gopalakrishnan Ramalingam, Jerrold A. Floro, and Petra Reinke[a)]

*Department of Materials Science and Engineering, University of Virginia, Charlottesville VA 22904*

**ABSTRACT**

The annealing of sub-critical Ge wetting layers (WL<3.5 ML) initiates the formation of 3D nanostructures, whose shape and orientation is determined by the WL thickness and thus directly related to the strain energy. The emergence of these nanostructures, hillocks and pre-quantum dots, is studied by scanning tunneling microscopy. A wetting layer deposited at 350 ºC is initially rough on the nanometer length-scale and undergoes a progressive transformation and smoothening until for T>460 ºC vacancy lines and the 2xn reconstruction are observed. The metastable Ge wetting layer (WL) then collapses to form 3D nanostructures whose morphology is controlled by the WL thickness: firstly, hillocks, with a wedding cake-type structure where the step edges run parallel to the <110> direction, are formed for thin wetting layers, while {105}-faceted structures, so-called pre-quantum dots (p-QDs), are formed from thicker layers. The wetting layer thickness and thus the misfit strain energy controls the type of structure. The crossover thickness between hillock and p-QDs regime is between 1.6 and 2.1 ML. The hillocks have larger lateral dimensions and volumes than p-QDs, and the p-QDs are exceptionally small quantum dots with a lower limit of 10 nm in width. Our work opens a new pathway to the control of nanostructure morphology and size in the elastically strained Ge/Si system.

[a)] Author to whom correspondence should be addressed. Electronic mail: pr6e@virginia.edu

# I. INTRODUCTION

The quantum effects in strain-induced Ge quantum dots (QDs) result in novel properties that make them candidates for use in photonic and electronic devices and solid-state quantum computing.[1–5] Ge deposition on Si(001) has been studied extensively and QD growth occurs from an initially flat Ge wetting layer (WL) which roughens as its thickness increases during continued deposition. Once the WL reaches a critical thickness, which corresponds to a critical strain energy, the formation of pyramidal quantum dots is observed and these transform, with continued Ge deposition, into large domes and superdomes.[4–8] The synthesis of ordered 2D and 3D arrays of Ge QDs has been achieved through the use pre-patterned and vicinal substrates where kinetic step bunching can induce an array of nano-patterns,[9] or seeding of nucleation centers by intentional introduction of impurities on a growth surface.[10]

The advances in understanding and manipulation of Ge QD nucleation and growth are contrasted by only few studies which investigate WLs with sub-critical thickness. Zhang et al.[11] recently demonstrated that monolithic Ge-nanowires can be synthesized by annealing of Ge-WLs with sub-critical thicknesses and their work renewed our interest in this growth regime. The metastable Ge WL was deposited at 560 °C and transformed into monolithic Ge wires upon annealing at a slightly lower temperature of 550 °C.[11] The monolithic nanowires, are {105}-faceted and only 2-3 unit cells in height,[11,12] and formed in a very narrow parameter space defined by the annealing time, temperature and WL thickness. Prolonged annealing leads to bunched wire structures.[13] Zhang et al.[14,15] also showed that the Ge wires could be used as templates to grow other novel Ge nanostructures such as dumbbells, dashes and dot-chains. These studies reignited the discussion about the equilibrium shapes of {105}-faceted Ge islands (pyramids, huts and wires).[12] Daruka et al.[12] re-examined the energetic contributions to the thermodynamic treatment of WL, wire and QD formation, and includes a new assessment of the excess surface energy (ESE) in which edge energies constitute a driving force for shape selection.

These observations have prompted us to widen the parameter space and study the response to annealing of WLs of variable thicknesses as a pathway to the formation of a wider range of nanostructures. Our work extends over a relatively large parameter space defined by the WL thickness (1.2 to 3.5 ML) and annealing temperature range (450 °C to 600 °C) to capture specifically the impact of WL thickness and consequently strain energy on roughening and relaxation of the WL into nanostructures. We have investigated wetting layers in the thickness range of 1.2-3.5 ML, which is below the critical WL thickness required to nucleate QDs at the growth temperatures used in this work.[16]

## II. EXPERIMENTAL SECTION

The experiments were performed in an Omicron Nanotechnology Variable Temperature SPM system under ultrahigh vacuum (UHV) conditions. The base pressure was ≤ 3×10$^{-10}$ mbar and the Si(100) samples were B-doped with resistivity between 0.01-0.02 Ω-cm. The sample was annealed overnight at 450 °C and thermally cleaned by flashing repeatedly to 1200-1300 °C to obtain a clean 2×1 reconstruction.[17,18] Ge (GoodFellow, 99.99% purity) was deposited from a Veeco effusion cell at a rate of 0.25 ML/min. Imaging was performed at room temperature in constant current mode with W tips prepared by electrochemical etching. Empty and filled states are imaged with a bias voltage of + 1.5 V and -1.5 V and a tunneling current of 0.03 nA. The temperature calibration procedure has been described by Nolph et al.[19] and the error in our temperature measurement is +/- 25 °C on account of the small sample size (5×1 mm$^2$) which prohibits reliable pyrometer measurements. Images were analyzed with Gwyddion, an open-source software for SPM data analysis.[20]

## III. RESULTS

Ge layers, with a thickness of 1.6 ML to 3.5 ML, were deposited at temperatures from 540-550 °C and subsequently annealed below the growth temperature for varying durations. Additionally, a 1.2 ML thick WL was deposited at 360 °C and annealed in multiple steps in the temperature range of 360-540 °C. The entire parameter space explored in this study is pictorially represented in a 3D plot of anneal temperature, annealing time and WL thickness in Figure 1(a). The samples are labeled by the initial WL thickness before annealing and is represented as S[WL-thickness]. A reference QD growth sample, $S_{ref}$, (3.8 ML at 470 °C) is used to calibrate the sample temperature and deposition rates and showed the characteristic hut-shaped QDs expected for a WL thickness exceeding the critical thickness for QD nucleation. Two distinct 3D structures are obtained when subcritical Ge wetting layers are annealed: (i) hillocks, shown in Figure 2-c and 2-d, with a wedding-cake type structure composed of stacked Ge(100) layers with narrow terraces, and (ii) pre-quantum dots (p-QDs), shown in Figure 3, which are {105}-faceted structures rotated by 45° compared to hillocks and similar to the conventional growth quantum dots.[5,16,21] The pre-quantum dots are so named because we are operating in a regime below the critical thickness required to nucleate quantum dots yet they already exhibit the {105} facet. In S[3.5], the terrace structure of the initial Si surface is reminiscent of a 4-6° miscut wafer and the corresponding p-QDs obtained in this experiment are triangular in shape as shown in Figure S1 (see supplementary information[22]) consistent with triangular QDs observed by Persichetti et al.[23] on miscut Si wafers.

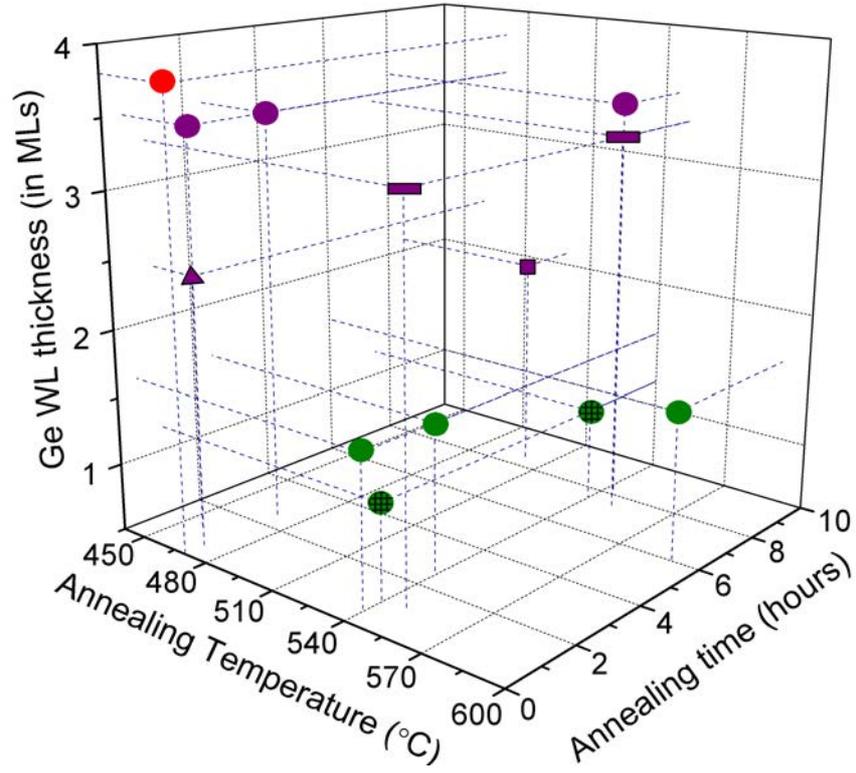

**Figure 1** (a) 3D parameter space map of all the samples prepared in the current work. The parameters that yield hillocks and p-QDs (pre-quantum dots) are color coded as green and violet respectively. The red circle in (a) is the reference QD sample. Each symbol in the map represents a sequential annealing experiment and is summarized in Table I.

**Table I.** Sample index with WL growth temperature and annealing temperatures used in this study.

| Sample Index | Growth Temperature (°C) | Annealing Temperature ( in °C) and Time | Structures Formed |
|---|---|---|---|
| S[1.2] | 360 | 405 (1h), 460 (0.75h), 540 (1h + 6.5h) | Hillocks |
| S[1.6] | 540 | 540 (0.5h + 2.5h), 590 (6h), 650 (2h) | Hillocks |
| S[2.1] | 550 | 500 (8.75h) | p-QDs |
| S[2.4] | 400 | 470 (1h) | p-QDs |
| S[3.3] | 570 | 540 (1h + 7h), 525 (1.5h) | p-QDs |
| S[3.5] | 475 | 470 (0.5h + 2h), 545 (8h) | p-QDs |
| $S_{ref}$ | 470 | - | Normal QDs |

The evolution of sample S[1.2], which was deposited at a low temperature of 360 °C, during sequential annealing from 400 °C to 540 °C is shown in Figure 2. The first layer of Ge completely wets the Si(100) surface owing to its low surface energy.[24,25] However, the Ge deposition at low temperatures results in a rough WL surface with many small islands 2-3 MLs in height due to kinetic barriers to diffusion as shown in the inset in Figure 2-a. Annealing for 1 hour at 405 °C does not significantly modify the surface morphology (as seen in Figure 2-a) indicative of a lack of long-range atomic diffusion which precludes the formation of a smooth layer. Dimer vacancy lines (DVLs), which are dimer vacancies running across the width of a terrace leading to a 2xn reconstruction, are not yet observed and vacancies are not yet mobile at these low temperatures. Annealing S[1.2] at 460 °C for one hour leads to significant reorganization of the surface and results in a smoother terrace structure (Figure 2-b) which is consistent with the WL structure seen during QD deposition.[26,27] The DVLs run across the entire width of a terrace resulting in the equilibrium 2x8 reconstruction for this system.

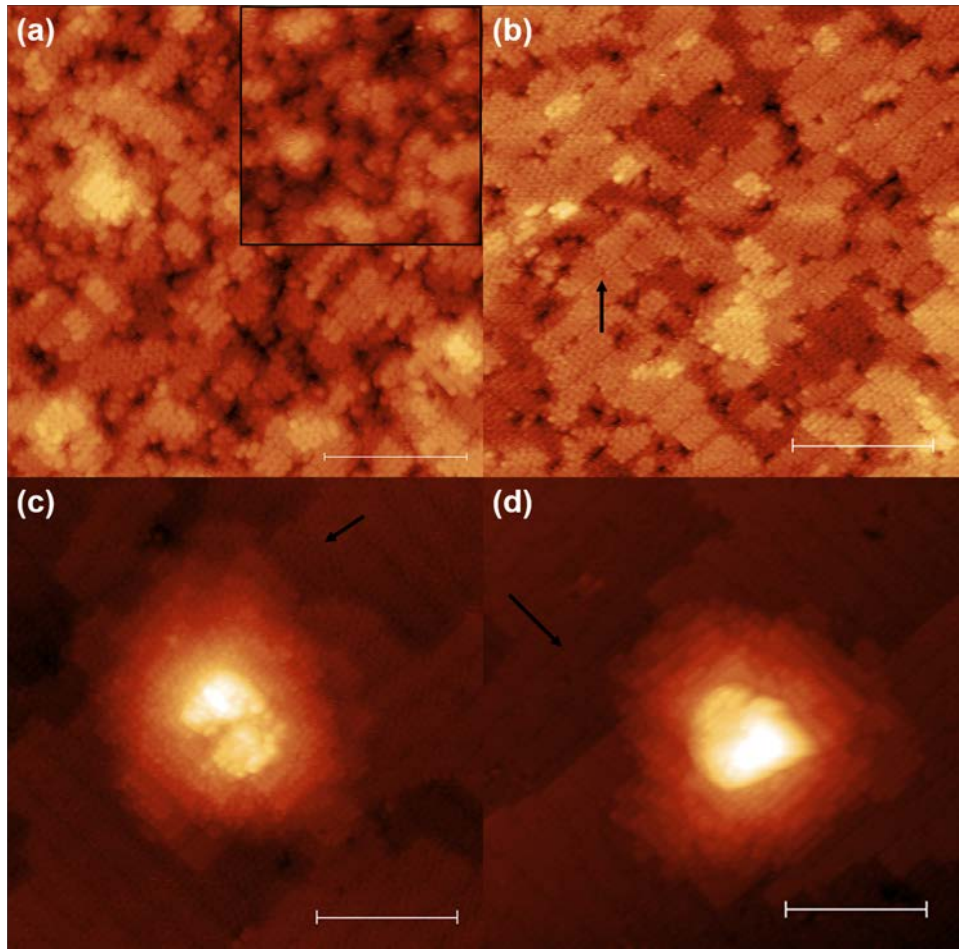

**Figure** 2. Evolution of S[1.2] when annealed at (a) 405 °C for 1 hour and the inset shows the as deposited surface (at 360 °C) with the same magnification, (b) 460 °C for 45 minutes, (c) 540 °C for 1 hour, and (d) 540 °C for 7.5 hours. Scale bar is 15 nm. Black arrows in (b), (c) and (d) point to DVLs on the surface.

Continued annealing of S[1.2] at 540 °C for 1 hour results in the formation of hillock structures which have not been observed previously in the Ge/Si system. Hillocks are characterized by a "wedding-cake type" stacked structure made up of Ge(100) terrace segments as seen in Figure 2(c) and 2(d). The hillock captured in Figure 2(c) and 2(d) are 1.4 and 3.4 nm high after cumulative anneal times of 1 and 7.5 hours and they make angles of 6.67±0.86° and 6.25±0.86° with the (100) plane (measured along <110>). This corresponds to an average terminating plane defined by the (001) step edges which is parallel to the

{$\bar{1}$ 1 12} plane. Annealing sample S[1.6] at temperatures and times indicated in Figure 1 and Table 1 yielded hillocks with identical crystallographic orientation but larger in size. The edge configuration in hillocks is determined by the goal to minimize the step length and the terrace width is controlled by a balance between the step formation energy and the repulsive interactions between the steps[28,29] which leads to local emergence of the stacking parallel to the {$\bar{1}$1 12} plane. The corresponding facet is assigned based on the angle between two adjacent steps[30] and if the step edge configuration is uniform, the overall angle the hillocks make with the (100) plane can be used. Based on the angle of 6.25±0.86° which the S[1.2] hillocks make with the surface, the hillock terminating plane facet can be assigned a Miller index of {$\bar{1}$1 12}.

Ge WLs with thickness greater than 2.0 ML evolve differently during annealing and form pre-quantum dots (p-QDs) which are characterized by the presence of {105} facets. The data points in purple in Figure 1 correspond to those samples which yield p-QDs. The transformation of the flat wetting layer into the faceted p-QDs as observed in the sample S[3.3] is shown in Figure 3 and representative of all sample where p-QDs are observed. The starting wetting layer surface shown in Figure 3(a) corresponds to the MxN reconstruction described in literature.[6,31,32] When annealed at 540 °C for 1 hour, small structures are nucleated on the surface some of which are partially {105}-faceted as seen in Figure 3(b). Prolonged annealing (+6 h at 540 °C) results in an increased number of fully faceted p-QDs (Figure 3c). The facet angle of 11.54±0.58° with the (100) surface confirms the {105} facet as known in Ge QDs. Note that not all the structures are faceted after extended annealing and a few partially-faceted and unfaceted structures are still present on the surface. High resolution images of the p-QDs and conventional QDs are presented in Figure S2. All 3D nanostructures have an amorphous mound at the apex similar to the observations of Sanduijav et al.[33] when the Ge/Si(100) samples were annealed between sequential Ge deposition cycles. They attributed these structures to growth defects resulting from extended storage of sample in ultrahigh vacuum (UHV) but we observe them in the annealing of freshly deposited samples. The discussion of amorphous apex structures is continued later once all datasets have been presented.

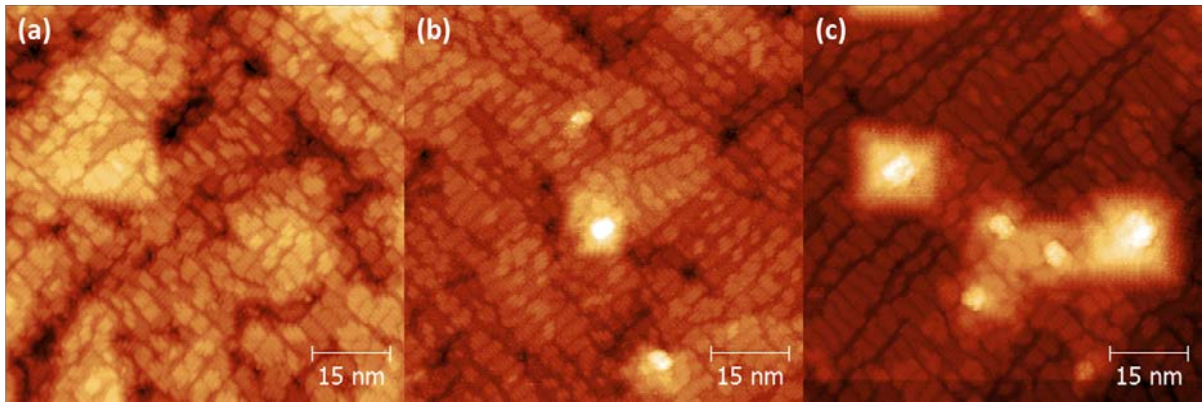

**Figure 3.** Evolution of 3.3 ML thick Ge wetting layer: (a) as deposited at 570 °C and after annealing at 540 °C for (b) 1 hour and (c) 7 hours. Partially and fully formed {105} facets on p-QDs are seen in (b) and (c).

One of the key parameters distinguishing hillocks and p-QDs is the angle made by the islands with the Ge(001) surface and this is expressed in Figure 4(a) as a function of the initial WL thickness. The hillocks make an average angle of 6.25±0.86° with the surface (measured along <110>) while the p-QDs across all experiments make an average angle of 11.18±0.61° (measured along <100>) in agreement with a {105} facet. Another striking observation is that the p-QDs are bounded by <100> directions, similar to conventional Ge huts and pyramids,[16,27,34] while the hillocks are rotated by 45° and are bounded by <110> directions (for reference, dimer rows and DVLs are oriented along <110>). The difference in orientation between hillocks and pre-quantum dots is clearly illustrated by STM images of structures from S[1.2] and S[3.3] (Figures 2 and 3) as well as in the insets in Figure 4(a). Poor imaging encountered during measurement of samples S[1.6] and S[2.4] precluded sufficiently precise measurements of facet angles.

Figures 4 and 5 illustrate the transition between the two types of nanostructures - hillocks and p-QDs. We observe in Figure 4(a) that hillocks only form at low coverages and that crossover from hillock formation to p-QDs occurs at a WL thickness between 1.6 ML and 2.1 ML, which can be related to the increased strain energy. The lower temperature limit for nanostructure formation is 460 ºC, where only transformations of the WL surface structure are observed. In addition, the first layer of Ge is conserved owing to its exceptional stability compared to the bare on Si(100)-2×1 surface.[25] We can now construct a stability diagram as shown in Figure 4(b) which predicts the type of structure that will be obtained for any given pair of WL thickness and annealing temperature ([WL, $T_{anneal}$]). For example, a combination of [3 ML, 425 °C] will result in reorganization of the WL with no 3D nanostructures while [3 ML, 500 °C] will yield {105}-faceted p-QDs. The initial WL thickness is the key variable that controls the type of nanostructure.

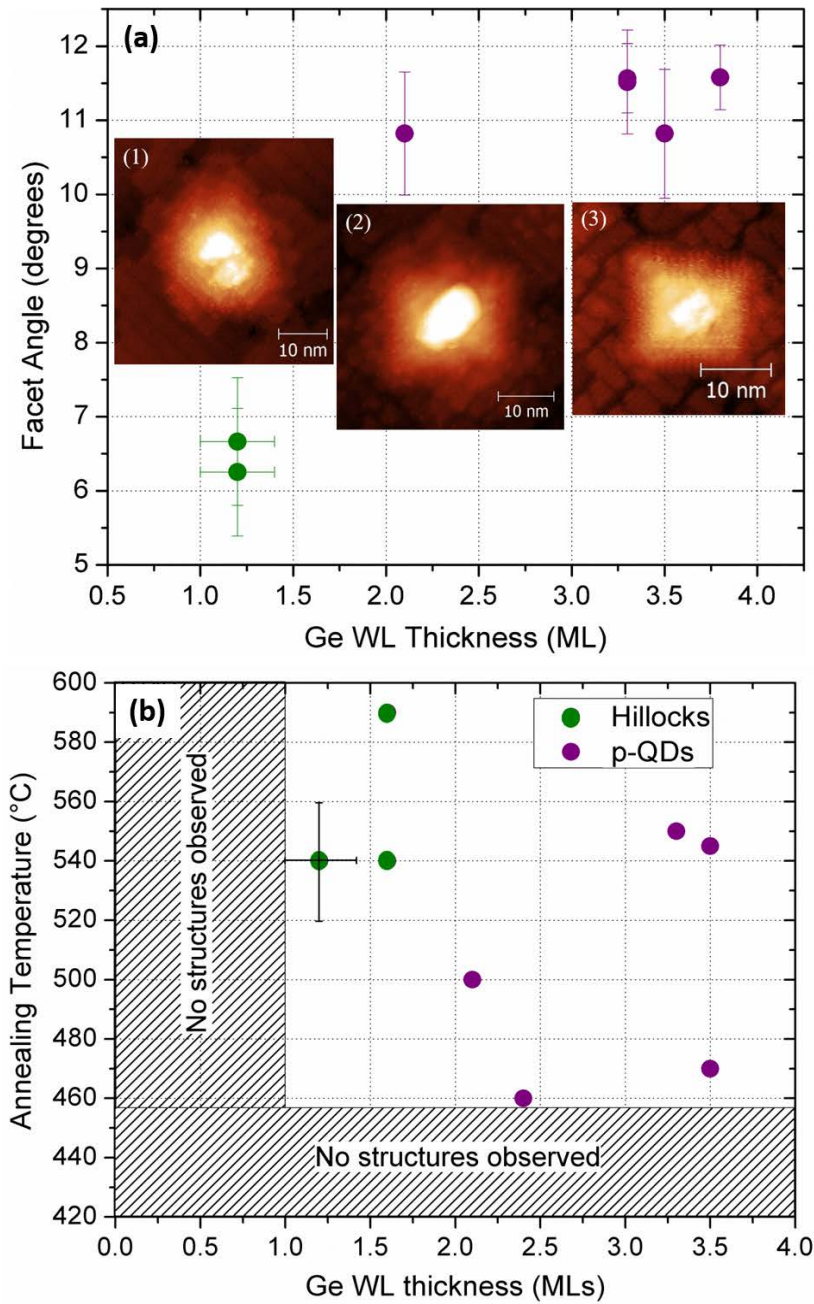

**Figure 4.** (a) Distribution of facet angles as a function of WL thickness with representative STM images of the structures obtained from S[1.2] (inset-1), S[2.1] (inset-2) and S[3.3] (inset-3). (b) WL thickness-annealing temperature map showing the minimum WL thickness and temperature required to nucleate hillocks and p-QDs. The error in our measurements is indicated for the data point corresponding to S[1.2].

The morphology of hillocks and p-QDs are summarized in Figure 5, which shows volume, and lateral dimensions as a function of WL thickness. Hillocks have a considerably larger volume compared to the p-QDs and the island number density follows the reverse trend where the number density of p-QDs is significantly higher than that of hillocks. The hillocks obtained after annealing S[1.2] at 540 °C for 1 hour have an average volume of 515±51 nm$^3$ ($\approx$2.27×10$^4$ atoms) and a density of 66 hillocks per $\mu$m$^2$. Prolonged annealing (540 °C for an additional 6.5 hours) increased in the density to 96 hillocks per $\mu$m$^2$ with an average volume of 655±75 nm$^3$ ($\approx$2.89×10$^4$ atoms). For sample S[3.3] discussed earlier (Figure 3), the p-QDs have an average volume of 43.13 nm$^3$ ($\approx$1903 atoms) with an island density of 637 p-QDs per $\mu$m$^2$ after annealing at 540 °C for 1 hour and it increases to 133.73 nm$^3$ ($\approx$5900 atoms) and 925 p-QDs per $\mu$m$^2$ after the subsequent 6 hour anneal. The absence of the steeper facets such as {113} and {15 3 23} that populate Ge domes even after prolonged annealing indicates that the misfit strain energy in the system is insufficient to compensate the energy cost for the formation of these facets while new islands with {105} facets are continually added as evidenced by the increase in p-QD density. S[2.1] shows an unusually low number density compared to other p-QDs which is likely due to the WL thickness being just above the crossover thickness and the insufficient driving force for the nucleation of more p-QDs. The presence of the crossover is a strong indication that the formation of hillocks and p-QDs follow fundamentally different mechanisms.

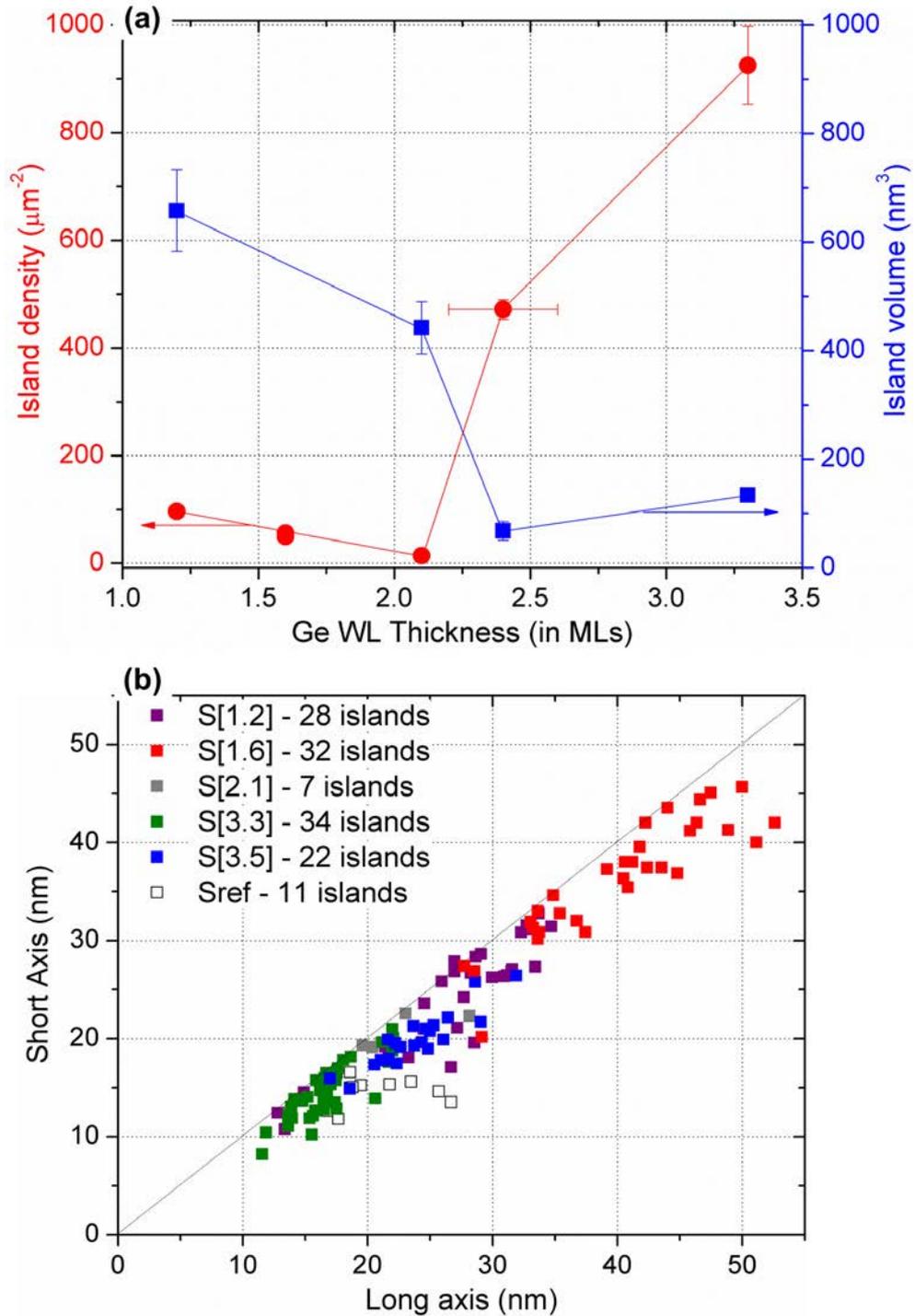

**Figure 5.** (a) Island volume and number density of hillocks and p-QDs as a function of Ge WL thickness. (b) Scatter plot of lateral dimensions of all hillocks, p-QDs, and the reference QD sample. The number in parenthesis refers to the wetting layer thickness.

The lateral dimensions of all the islands, including reference quantum dots from $S_{ref}$, are expressed in Figure 5(b). This plot shows clearly that the faceted p-QDs, on average, have smaller in-plane dimensions (length and width) compared to hillocks: for example, p-QDs from sample S[3.3] have final lateral dimensions of 16.5 and 16.0 nm (with standard deviation of 2.2 and 2.6 nm respectively) while the dimensions of hillocks obtained in sample S[1.6] are more than twice as large with 40.3 and 36.7 nm (standard deviation of 4.3 and 4.1 nm respectively). Both groups prefer a nearly square profile, and some p-QDs present a very small base length of 10 nm, which is only very rarely observed in Stranksi-Krastanov growth of QDs. Hillocks grow faster compared to p-QDs and the evolution of the structures when annealed is summarized in Figure S3. This figure shows a complete set of size distributions for hillocks (S[1.6]) and p-QDs (S[3.3]) visualizing the significantly slower growth of p-QDs. The slow growth rate of {105} facets[35] can be attributed to kinetic limitations in the nucleation of a new layer on the {105} facet. Figure S3 also confirms that the structures observed are not metastable fluctuations which dissolve beyond a threshold barrier and that they are, in fact, growing structures.

Mass balance calculations were performed to determine the amount of Ge consumed in the formation of hillocks and p-QDs. Among all the samples prepared, only a fraction of the available Ge in the WL is consumed and the maximum amount of Ge consumed was 0.44 ML in case of S[1.6] at the end of the final annealing step. For example, in sample S[3.3], only 0.07±0.03 ML of Ge was incorporated in the p-QDs after the first annealing step (540 °C for 1 hour). After the subsequent anneal for 7 hours, an additional 0.26±0.08 ML of Ge was consumed in the formation of p-QDs with 2.99±0.08 ML of wetting layer remaining. In case of hillock formation in S[1.2], 0.07±0.01 ML and 0.17±0.02 ML of Ge are consumed after annealing for 1 hour and 6.5 hours, respectively, which is slightly lower than for S[3.3] using a similar annealing cycle. These results confirm that the formation of three-dimensional islands is not limited by the supply of Ge atoms. The first Ge-layer in contact with the Si-substrate is never consumed, which is consistent with existing literature[25] which shows that a monolayer of Ge on Si(001) has a lower surface energy than a bare silicon surface. A "wetting potential" has often been invoked to explain the stabilization of the wetting layer to thicknesses significantly in excess of 1 ML. However, our results could imply that this stabilization is largely kinetic in origin, rather than being due to a true energetic interaction.

**IV. DISCUSSION**

The fundamental aspects of Ge quantum dot growth on Si(100) have been studied intensely and while many aspects of QD growth are well-understood, the multiple factors controlling island shape remain open to discussion.[4,12,16,27,31] The elongation of a pyramidal Ge QD into huts and wires has traditionally been attributed to kinetic factors but Daruka et al.[12] used basic thermodynamic parameters such as strain energy, surface energy, and edge energy to determine the stability of pyramidal, hut and wire structures under equilibrium conditions. In heteroepitaxial growth of faceted islands, the excess surface energy (ESE) is defined as the difference between the total surface energy of the system in the presence of islands and without islands (strained wetting layer).[12] By taking into account the edge and kink energies and the surface energy of the strained {105} facet,[36] it has been shown that ESE is negative for conventional pyramidal and hut quantum dot structures as well as for elongated wire structures on a Ge WL surface,[11,37] whose synthesis provided the initial motivation for our study. The delicate energy balance between surface energy and strain energy stored in the Ge-WL determines whether QDs or wires are formed assuming the temperature is sufficient to allow for the expression of the equilibrium structure. Earlier work by Tersoff et al.[21] focused on the initial formation of QDs, and suggests a process where small, unfaceted mounds transform into faceted QDs based on several STM studies.[4,5,16,38]

The formation of hillocks can be understood in the framework of the evolution of the Ge-WL surface as a function of temperature illustrated in Figure 2: the initial surface is rough on a very short length-scale and in response to annealing, the surface roughness decreases, the step edges thaw, and the dimer vacancies become mobile and condense in dimer vacancy lines. A further increase in the annealing temperature triggers the formation of hillocks with a roughly square footprint and wedding cake like structure. The sides of the square are always perpendicular to the direction of step bunching on Si and Ge(001) surfaces, which is the <110> direction, and their slope is at a near constant angle, which can be interpreted as the saturation angle representative of the stable configuration at this temperature. The hillock formation is observed below the roughening transition temperature[39,40] where the free energy of step formation vanishes, and can be better described as localized step bunching. The inverse Ehrlich-Schwöbel barrier is sufficiently small to allow for an upward adatom flux thus destabilizing the terrace structure and leading to reduction in the local terrace width at the hillock but increasing the terrace size in between hillocks. The magnitude of the inverse Ehrlich-Schwöbel barrier is likely influenced by the line tension and edge structure in the hillocks, and the compressive stress in the Ge-WL.[41] The wedding-cake type structure of hillocks is reminiscent of 3D structures that result from massive step bunching on the surface; considerable step motion is expected at the annealing temperatures used (>540 °C) and we propose this as a possible mechanism for the growth of hillocks. Prior work on the formation of QDs in a conventional growth process indicates that small surface mounds on the WL (pre-pyramids) initiate the formation of the {105} faceted QDs.[5] We propose that the unfaceted and partially faceted islands in Figure 3(b) are identical to the mounds observed by Vailionis et al.[5] and are precursors for both hillocks as well as p-QDs (as seen in the sequence in Figure 3).

The next question that arises from the current work is to determine the parameter(s) that controls the type of structure. The type of structure formed (hillock versus p-QDs) is dictated by the intricate balance between the various energies involved in the nucleation and growth process: (a) surface energy of the initial metastable WL and the final configuration with hillocks/p-QDs, (b) energy cost for formation of faceted p-QDs or step-edge creation in case of hillocks (dangling bonds at edges), and (c) release of misfit strain energy, an intrinsic function of WL thickness, which follows the formation of 3D islands. Under the conditions used here, the cross-over between hillock and p-QD formation is relatively abrupt and is tied to the WL thickness which is proportional to the magnitude of the strain energy stored in the WL. Figure 4 shows the abrupt transition between hillocks and p-QDs with a crossover thickness between 1.6 and 2.1 ML. We did not observe any direct transformations from hillocks to p-QD - either one or the other is formed and then continues to grow in size during the continued annealing process. In this sense, our results for annealing Ge wetting layers differ from those of Vailionis et al.[5] and Tersoff et al.[21] where a direct transition between the two shapes occurs during continuous deposition.

The formation of hillocks/p-QDs is likely initiated by surface or subsurface defects and adventitious impurities, which act as a sink for mobile Ge species on the surface.[40] Some intermixing is also expected in this system due to the annealing at these temperatures as has been observed by Zhang et al.[11] The amorphous structure at the apex of QDs has been observed previously by Vailionis et al.[5] and Sanduijav et al.[33] during the conventional fabrication of Ge QDs and in case of the latter, it was attributed to the presence of impurities. In our experiments, the amorphous apex structure is confined to the hillock/p-QD regime as shown in Figures 2 and 3 and is not present in the QDs grown in the reference experiment $S_{ref}$ – this eliminates the Ge source as the origin of contamination. A control experiment where a Si(001)-(2x1) substrate was annealed at 600 °C for 2 hours did not show any surface structures like hillocks, step bunching or reconstructions, which are typically associated with carbon contamination.[40] While the presence of impurities can never be completely excluded, in light of the results presented here, we favor the development of an inherent surface instability as the origin of hillock and p-QD formation.

In conclusion, we have demonstrated the conversion of the metastable, sub-critical Ge wetting layers grown on a Si(001)-2×1 surface into two distinct types of 3D nanostructures by a post-growth annealing process. Hillocks have a wedding-cake type stacked structure while p-QDs are {105}-faceted structures similar to conventionally grown QDs. The WL thickness (correlated with misfit strain energy) is the key parameter that controls the type of structure obtained and a stability map is developed which predicts the type of nanostructure which will emerge. The minimum annealing temperature for the nucleation of any nanostructure is 460 °C and the crossover WL thickness above which {105}-faceting is stable and p-QDs form is between 1.6 and 2.1 ML. These p-QDs can be exceptionally small and reach lateral dimensions down to 10 nm.

Our results demonstrate the possibility of obtaining quantum dots at WL thicknesses below the nominal equilibrium critical thickness used in current literature to demarcate the thickness where {105} faceted QDs emerge. This could suggest that the true equilibrium wetting layer thickness is significantly smaller, perhaps only a single monolayer of Ge. However, more work is needed to assess this. Indeed, if a growth experiment with very low Ge deposition rates were to be performed for example: 2 hours to deposit 3 ML of Ge at 550 °C (critical thickness at 550 °C is ≈4.5 ML according to the literature on growth experiments), it might yield fully-formed {105}-faceted QDs despite not crossing the critical thickness as it is currently defined. This would imply that the critical thickness for forming Ge QDs depends on the growth rate. It also indicates that the concept of critical thickness has to be expanded and include pathway along the transformation of 3D structures, including formation of hillocks or p-QDs described here.

## V. ACKNOWLEDGEMENTS

The authors acknowledge the support of this work by the National Science Foundation (NSF) through award number DMR-0907234 by the Division of Materials Research (Electronic and Photonic Materials).

# 3D Nanostructures on Ge/Si(100) Wetting Layers: Hillocks and Pre-Quantum Dots

Gopalakrishnan Ramalingam, Jerrold A. Floro, and Petra Reinke[a]

*Department of Materials Science and Engineering, University of Virginia, Charlottesville VA 22904*

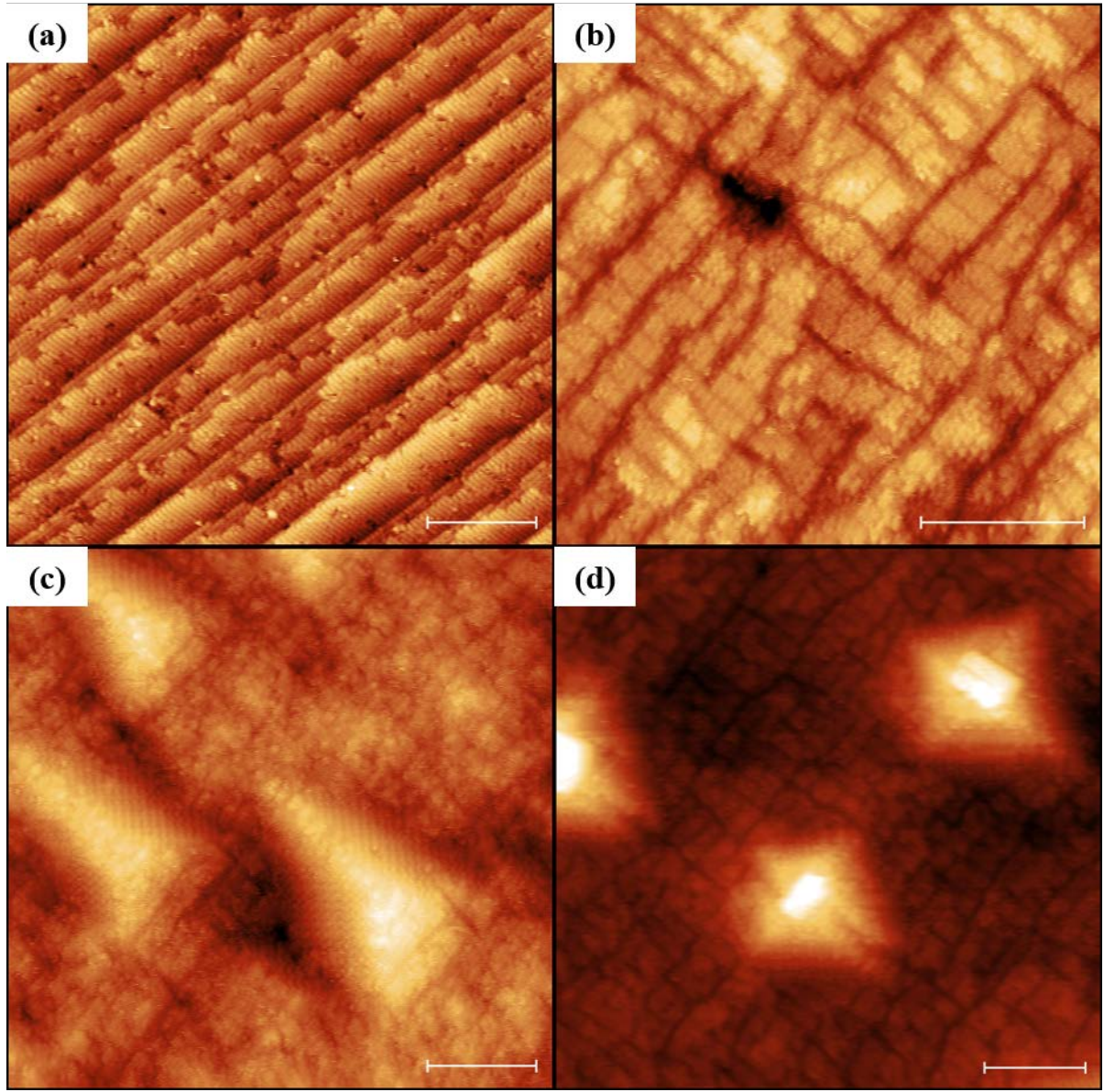

**Figure S1.** STM images showing the evolution of S[3.5] during post-growth annealing: (a) initial Si surface with the terrace structure similar to 4° miscut wafer caused by mechanical strain exerted on the sample during the high temperature anneal for substrate cleaning,[1] (b) Ge WL deposited at 475 °C with the MxN reconstruction, (c) triangular p-QDs[1–3] showing {105} facets after annealing at 470 °C for 0.5 hours and elongated along <110>, and (d) after an additional annealing step at 545 °C for 8 hours showing almost-square p-QDs (with residual pointed ends along <110>). Scale bar is 15 nm.

[a] Author to whom correspondence should be addressed. Electronic mail: pr6e@virginia.edu

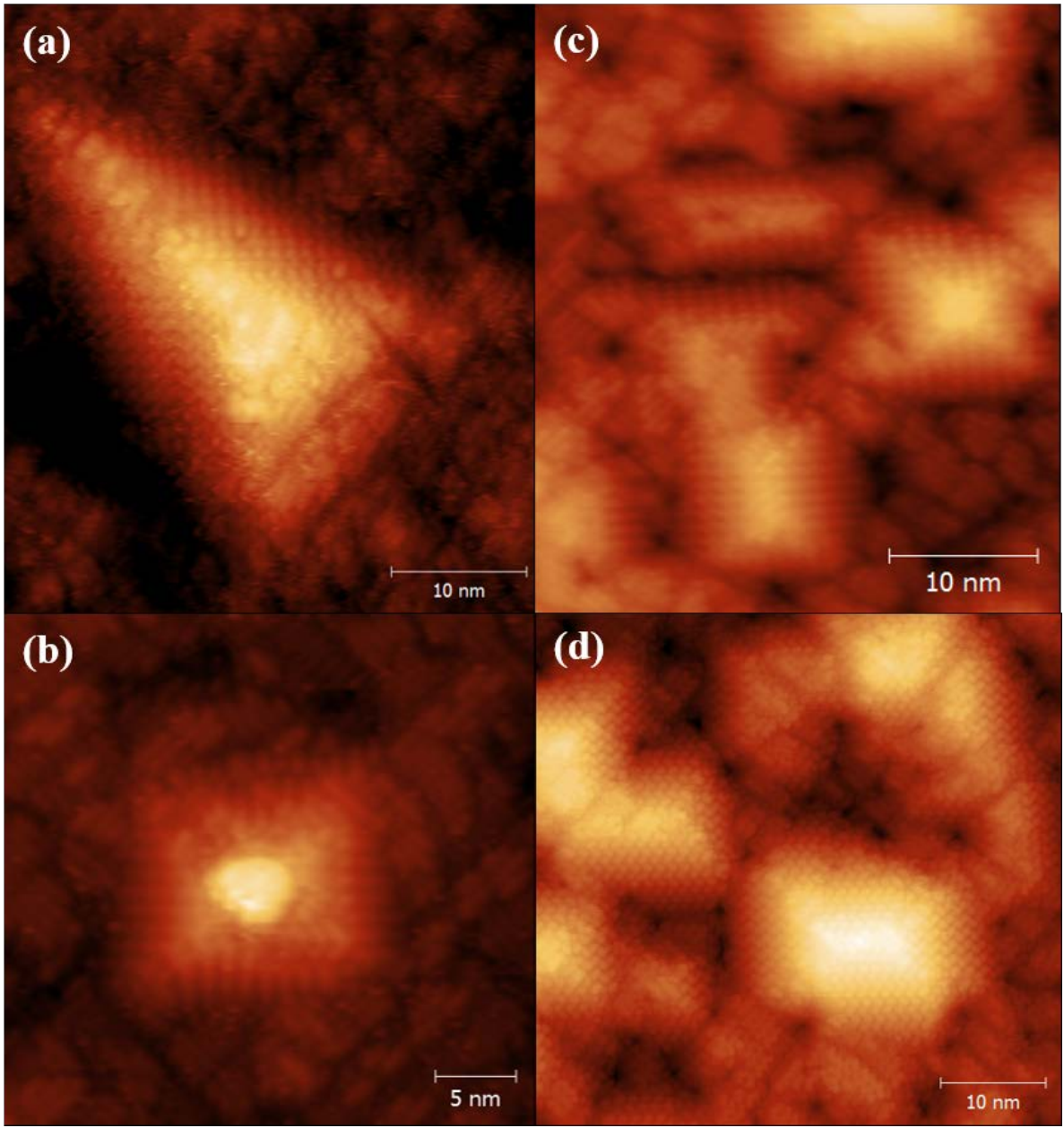

**Figure S2.** High resolution STM images of pre-quantum dots (p-QDs) and conventional quantum dots: (a) and (b) show images of p-QD from S[3.5], filled states, and S[3.3], empty states. (c) and (d) show empty and filled states images of conventional QDs from $S_{ref}$. The reconstructed {105} surface is clearly visible and consistent with literature.[4–6] Empty states images were recorded at +2.0 V, 0.1 nA and the filled states images were recorded at -2.0 V, 0.1 nA.

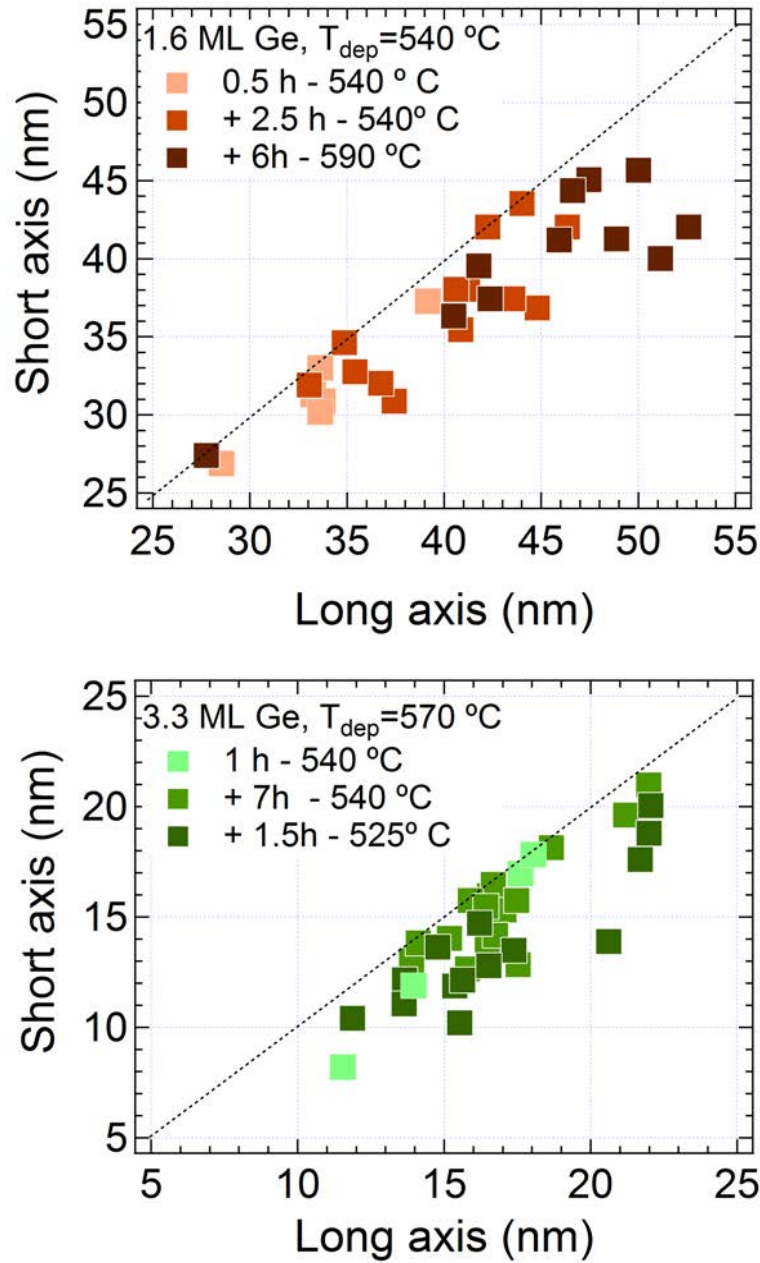

**Figure S3.** The evolution of (a) hillocks (from sample S[1.6]) and (b) p-QDs (from sample S[3.3]) with sequential annealing steps after Ge wetting layer deposition. The hillocks are, on average, larger than the corresponding p-QDs and also show increasing lateral dimension with multiple annealing steps. The size and shape distribution for all QDs are included in Figure 5 of the main text.